\documentclass[12pt]{iopart}
\usepackage{graphicx}
\usepackage{cite}
\usepackage{amsfonts}
\begin{document}

\title[Ageing in homogeneous systems at criticality]
      {Ageing in homogeneous systems at criticality}
\author{Haye Hinrichsen}
\address{Universit\"at W\"urzburg\\
	 Fakult\"at f\"ur Physik und Astronomie\\
         D-97074 W\"urzburg, Germany}
\ead{hinrichsen@physik.uni-wuerzburg.de}

\begin{abstract}
Ageing phenomena are observed in a large variety of dynamical systems exhibiting a slow relaxation from a non-equilibrium initial state. Ageing can be characterised in terms of the linear response $R(t,s)$ at time $t$ to a local perturbation at time $s<t$. Usually one distinguishes two dynamical regimes, namely, the quasi-stationary regime, where the response is translationally invariant in time, and the ageing regime, where this invariance is broken. In general these two regimes are separate in the sense that the two limits of \textit{(a)} taking $t,s\to\infty$ while keeping $t/s$ fixed, and \textit{(b)} taking $t,s\to\infty$ with fixed $t-s$, give different results. In recent years, ageing was also investigated in the context of \textit{homogeneous critical} systems such as the Glauber-Ising model and the contact process. Here we argue that, in contrast to a widespread believe, homogeneous critical systems do not have a separate quasi-stationary regime, hence the two limits do commute. Moreover, it is discussed under which conditions two particular exponents, denoted as $a$ and $a'$ in the literature, have to be identical.
\end{abstract}

\submitto{Journal of Statistical Mechanics: Theory and Experiment}


\parskip 2mm 

\section{Introduction}

In statistical physics a dynamical system prepared in a certain initial state is said to exhibit \textit{ageing} if its physical properties vary slowly with time~\cite{Young98}. Such a time dependence is typically observed in disordered systems such as spin glasses which memorise their initial state for very long time. As a hallmark of ageing, one observes that quantities such as two-point correlation functions are not translationally invariant in time. 

\begin{figure}[t]
\begin{center}
\includegraphics[width=120mm]{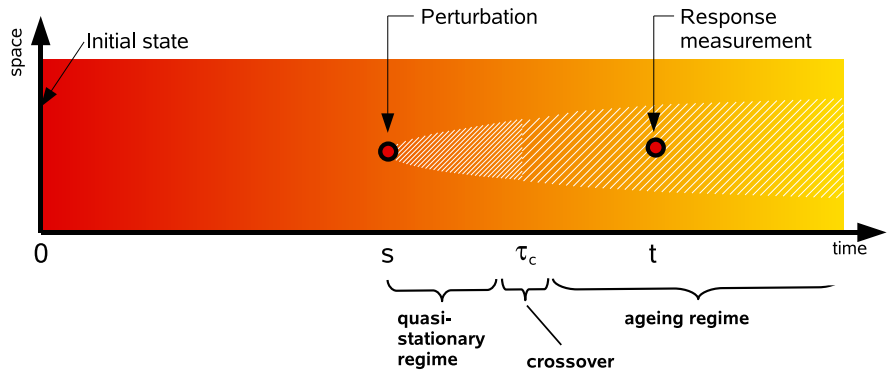}
\end{center}
\caption{\label{fig:scheme}\small
Cartoon of the autoresponse function $R(t,s)$ measured in a disordered system (see text).}
\end{figure}

Ageing can be characterised in various ways, e.g. by studying the autoresponse function $R(t,s)$. This function quantifies how a weak perturbation by an external field~$h$ at time $s$ exerted locally at position~$\vec{x}$ will change the averaged order parameter $\langle\phi(\vec{x},s)\rangle$ at some later time $t>s$:
\begin{equation}
R(t,s) = \left.\frac{\delta \langle \phi(\vec{x},t) \rangle}{\delta h(\vec{x},s)}\right|_{h(\vec{x})=0}.
\end{equation}
Fig.~\ref{fig:scheme} sketches how the autoresponse function is measured. At $t=0$ the system is quenched to a non-equilibrium initial state. As time evolves the system relaxes slowly towards the stationary state (without being able to reach it), accompanied by a gradual change of the averaged order parameter, as indicated by the changing background colour in the figure. After a certain waiting time $s$ a local external field is switched on for a short time, imposing a local change of the order parameter (e.g. a spin flip). As time proceeds this perturbation spreads in space and becomes less intense, as expressed schematically by the parabola-shaped region in the figure. The autoresponse function is just the average variation of the order parameter measured at time $t>s$ at the same position from where the perturbation originated.

In numerous studies of disordered systems it was observed that the temporal evolution after the perturbation can be divided into two different dynamical regimes (see e.g. Zippold {\it et al.}~\cite{ZippoldEtAl00}). Initially the influence of the external field dominates and causes a response which is essentially insensitive to the underlying slow relaxation of the entire system. In this so-called \textit{quasi-stationary regime} the response function is translationally invariant in time, i.e. it depends to leading order on $t-s$ but not on $s$:
\begin{equation}
\label{QuasiStationary}
R_{\rm st}(t,s)\;=\; R_{\rm st}(t-s).
\end{equation}
After some time, however, the response becomes weaker and begins to `feel' the relaxation of the system. In this so-called \textit{ageing regime} the response function $R_{\rm ag}(t,s)$ is no longer translational invariant in time, rather it depends on both parameters $t$ and $s$ in a non-trivial way. Both regimes are connected by a smooth crossover taking place at some typical time scale $\tau_c(s)>s$. Surprisingly, in disordered systems this crossover time was found to scale as~\cite{ZippoldEtAl00}
\begin{equation}
\label{Anomalous}
\tau_c(s)-s \,\sim \,s^\gamma
\end{equation}
with an exponent $0<\gamma<1$. In other words, the temporal size of the quasi-stationary regime grows with the waiting time $s$ but it grows slower than $s$ itself. 

\begin{figure}[t]
\centering\includegraphics[width=100mm]{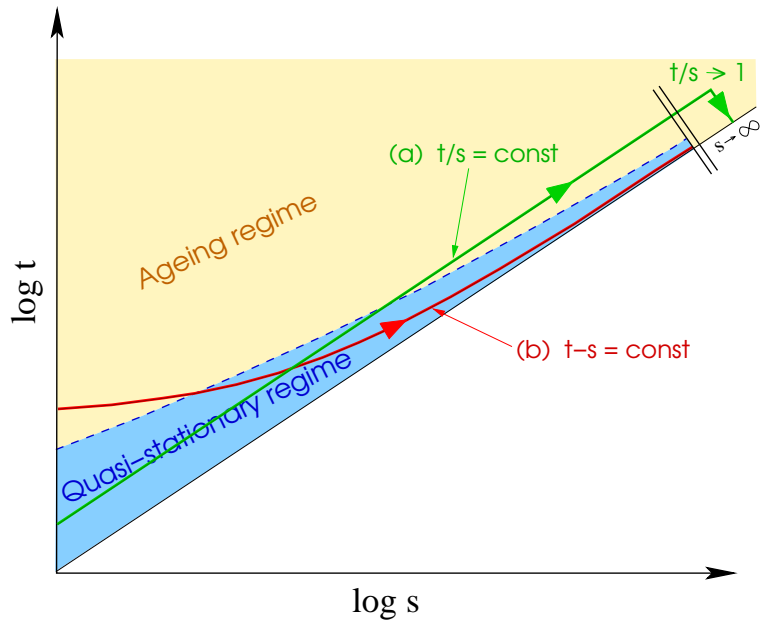}
\caption{\label{fig:climits}\footnotesize
Inequivalence of the limits (a) and (b). According to~\cite{ZippoldEtAl00} the response to a perturbation can be split into a quasi-stationary regime and an ageing regime, connected by a crossover at a typical time scale $\tau_c(s)>s$ (dashed blue line), where $\tau_c(s)-s\sim s^\gamma$ with $0<\gamma<1$ (here $\gamma=1/2$). The green and the red line symbolise typical paths along which the limits (a) and (b) are taken (see text).}
\end{figure}

\noindent
As noted by Zippold {\it et al.}, Eq.~(\ref{Anomalous}) has the non-trivial consequence that the two limits of
\begin{quote}
\begin{itemize}
\item[(a)] first taking $t,s \to \infty$ with $\xi=t/s$ fixed and \textit{then} sending $\xi \to 1$, and\vspace{3mm}
\item[(b)] sending $t,s \to \infty$ while keeping the difference $t-s$ fixed
\end{itemize}
\end{quote}
are generally different. This inequivalence is illustrated in Fig.~\ref{fig:climits}, where the two dynamical regimes are sketched in a double-logarithmic $t$-$s$-plane. The typical path according to limit~(a) is marked as a green line. As can be seen, this line runs parallel to the diagonal and enters the ageing regime after some time. When $s,t$ are taken to infinity (moving along the diagonal) the width of the quasi-stationary regime in the log-log representation shrinks to zero. Therefore, the second step of (a), the limit $t/s\to 1$, takes place entirely within the ageing regime and hence this limit is determined exclusively by the properties of the ageing regime. Contrarily, in the case of limit (b), where the difference $t-s$ is kept constant, the path (red line) eventually enters the quasi-stationary region, hence this limit will be determined by the properties of the quasi-stationary regime only. For this reason it is plausible that the two limits give generally different results.

The interest in ageing phenomena stems from the fact that the ageing regime displays scale-invariant features. This means that a change of time scales $s\to b s,\,t\to b t$ by a factor $b>0$ can be compensated by an appropriate rescaling of the response function $R\to b^{-1-a}R$, where $a$ is a certain exponent depending on the scaling dimension of the order parameter and the response field. Scale invariance in the ageing regime implies that the response function obeys the scaling form
\begin{equation}
\label{QuotientScalingForm}
R_{\mathrm{ag}}(t,s)  \sim s^{-1-a} f_R(t/s)\,,\qquad t\gg \tau_c(s)
\end{equation}
where $f_R$ is a scaling function. For large arguments it is found that the scaling function decays algebraically as
\begin{equation}
f_R(\xi) \sim \xi^{-\lambda_R/z}\,,\qquad\qquad\xi\to\infty
\end{equation}
where $\lambda_R$ is a new non-trivial exponent, the so-called ageing exponent, while $z$ denotes the dynamical exponent of the ageing regime. The scaling function and the exponents $a,\lambda_R,z$ are expected to be universal, i.e., they are determined by the symmetry properties of the model irrespective of its microscopic details.

In recent years the concept of ageing was also applied to \textit{homogeneous} critical systems with a continuous phase transition~\cite{HenkelPleimling03} such as the 2D Ising model~\cite{McCoyWu73} and directed percolation~\cite{Hinrichsen00,Odor04,Lubeck04} without quenched disorder. However, these models differ from previously studied systems such as spin glasses in so far as they have a much simpler scaling behaviour. The purpose of this contribution is to point out that in this case -- contrary to a widespread believe -- the two dynamical regimes are not separate but connected by an ordinary scale-invariant crossover, meaning that the two limits (a) and (b) give identical results.

The second issue addressed in this paper concerns a recently proposed theory of local scale invariance (LSI), which introduces an additional exponent $a'$. The arguments presented here imply that for the 2d Ising model and the contact process one has $a=a'$.

\section{Ageing in systems at criticality}

In what follows let us consider a stochastic process on an infinite lattice with the following properties:
\begin{enumerate}
	\item The model is homogeneous in space and time (no quenched disorder) and exhibits a continuous phase transition.

	\item The transition is characterised by \textit{simple scaling} (as opposed to multiscaling) involving three critical exponents, namely, the order parameter exponent $\beta$, the correlation length exponent $\nu=\nu_\perp$, and the dynamical exponent $z=\nu_\parallel/\nu_\perp$.

	\item The system has an autonomous response in the stationary state, i.e., the response function $R(t,s)$ at criticality is non-zero in the limit $s\to\infty$  with $t-s$ fixed.
\end{enumerate}
Examples of such models are the critical Ising model in two dimensions (2d) with standard heat bath dynamics as well as critical directed percolation in 1+1 dimensions. These models display asymptotic scaling laws which are well understood.\footnote{Note that any lattice model displays corrections to scaling caused by lattice effects on short distances and time scales (denoted as $\tau_{\rm micro}$ in the literature). Throughout this paper we assume that distances and time scales are large enough so that such lattice effects can be neglected.} Clearly, these scaling laws are valid in the stationary state at criticality and thus they will determine the scaling properties in the quasi-stationary regime. However, in contrast to disordered systems such as spin glasses, the \textit{same} scaling laws are also valid in the ageing regime $t\gg s$. As will be shown below, this implies that the scenario suggested by Zippold {\it et al.}~\cite{ZippoldEtAl00} does not hold for such systems. 

To see this let us assume the existence of two dynamical regimes separated by a crossover at a typical time scale $\tau_c(s)-s \,\sim \,s^\gamma$ with $\gamma <1$. In the quasi-stationary regime, where the response function $R_{\mathrm{st}}(t,s)=R_{\mathrm{st}}(t-s)$ is translational invariant in time, invariance under rescaling
\begin{equation}
\label{ScaleInvariance}
t \to \lambda t\,,\quad s \to \lambda s\,, \quad \vec x\to \lambda^{1/z} \vec x\,,\quad R \to \lambda^{-\beta/\nu z} R
\end{equation}
implies that $R(\lambda t-\lambda s)=\lambda^{-\beta/\nu z}R(t-s)$, leading to the well-known result
\begin{equation}
\label{StationaryDecay}
R_{\mathrm{st}}(t,s) \;\simeq\; r_0\, (t-s)^{-\beta/\nu z}\,. \qquad t\ll\tau_c(s).
\end{equation}
Note that for models with an autonomous response in the stationary state, the proportionality constant $r_0$ does not depend on $s$.

In the ageing regime, the system is still invariant under the \textit{same} type of scale transformation~(\ref{ScaleInvariance}) but translational invariance in time is broken, leading to the weaker scaling form
\begin{equation}
\label{NonStationaryDecay}
R_{\mathrm{ag}}(t,s) \;\simeq\; s^{-\beta/\nu z} f_R(t/s)\,, \qquad t\gg\tau_c(s).
\end{equation}
hence $a=\beta/\nu z-1$ in Eq.~(\ref{QuotientScalingForm}). This scaling form has been verified for various systems and is undisputed in the literature.

Both scaling forms are smoothly connected at the crossover $t=\tau_c(s)$, that is $R_{\mathrm{st}}(t,\tau_c(s))\simeq R_{\mathrm{ag}}(t,\tau_c(s))$, leading to the condition
\begin{equation}
\label{FixScalFunc}
r_0 f_R(\xi_c)\;\simeq\;(\xi_c-1)^{-\beta/\nu z}\,,
\end{equation}
where $\xi_c:=\tau_c(s)/s$. For $\gamma<1$ the argument $\xi_c$ varies with $s$, fixing the functional form of the scaling function $f_R$. However, this particular functional form would imply translational invariance $R_{\mathrm{ag}}(t,s)=R_{\mathrm{ag}}(t-s)$ in parts of the ageing regime, which is definitely not true. Therefore, the scenario suggested by Zippold \textit{et al.} does not hold for critical systems such as the 2d Ising model or directed percolation. The only way out is to set $\gamma=1$ so that Eq.~(\ref{FixScalFunc}) determines the scaling function only in a single point. This means that the crossover time $\tau_c(s) \sim s$ scales like any other time scale, which is quite natural in the context of second-order phase transitions. In the representation of Fig.~\ref{fig:climits} the ageing regime would be a strip of constant width. As can be seen, this immediately implies that the two limits (a) and (b) will give identical results.

\section{Test of Local Scale Invariance}

Recently a theory of Local Scale Invariance (LSI)~\cite{HenkelPleimling03,Henkel07} has been applied to various homogeneous many-particle systems with second-order phase transitions. This theory assumes that the linear response function formed from so-called quasi-primary fields transforms covariantly under the action of a group of local scale transformations. Like conformal invariance, this extended symmetry determines the functional form of scaling functions. In particular, LSI predicts that the scaling function $f_R(\xi)$ factorises into two power laws (see Eq.~(26) of Ref.~\cite{Henkel07})
\begin{equation}
f_R(\xi) \;=\; f_0\, \xi^{1-a'-\lambda_R/z}\,(\xi-1)^{-1-a'}.
\end{equation}
with an additional exponents $a'$ which may be different from $a$. However, knowing that the aforementioned limits commute, this scaling form would imply that for $t/s\to1^+$ the response function behaves as
\begin{equation}
R(t,s) \;\sim\; f_0\,\Big( t\,s^{a'-a} -s \Big)^{-1-a'}\,.
\end{equation}
Obviously, this behaviour is in contradiction with Eq.~(\ref{QuasiStationary}) unless $a'=a$ or $f_0=0$. 

Originally the theory of LSI was formulated for systems with translational invariance in time, which led to the constraint $a'=a$. The initial hope was that this theory is fundamental enough to describe a broad range of apparently unrelated models. However, later it became clear that the critical Ising model in 2d and the 1+1-dimensional contact process do not obey the predicted scaling forms for the autoresponse function~\cite{CalabreseGambassi02,Hinrichsen06,LippielloEtAl06}. As a way out, Henkel and collaborators generalised the theory, relaxing the symmetry by giving up translational invariance in time which allows $a$ and $a'$ to be different. Fitting numerical data of various critical models to the predicted response function they found a good agreement when~\cite{Henkel07}
\begin{small}
\begin{equation*}
a'-a \;=\; \left\{
\begin{array}{ll}
-0.187(20) & \mbox{ for the 2d Ising model with heat bath dynamics,} \\
-0.022(5)  & \mbox{ for the 3d Ising model with heat bath dynamics,} \\
+0.270(10) & \mbox{ for the 1+1-dimensional contact process.}        \\
\end{array} \right.
\end{equation*}
\end{small}
However, these models have an autonomous response ($f_0>0$), hence the above arguments require that $a'=a$. This suggests that the numerical estimates reported above have no deeper physical meaning; they merely result from the attempt to fit a wrong scaling function to numerical data within a limited window of $\xi$.

The proponents of generalised LSI refer to the circumstance that for the 1d Ising model one can prove that $a'-a=-1/2$. However, this model becomes critical at zero temperature with a non-fluctuating steady state so that it does not have an autonomous response in the limit $s\to\infty$, hence $f_0=0$. Therefore, it is no surprise that $a\neq a'$.

\section{Conclusions}

Ageing was originally introduced as a concept to describe the propagation of pertur\-bations in disordered systems. As a key property, such systems show two separate dynamical regimes, namely, a quasi-stationary one and the so-called ageing regime.

In homogeneous critical systems such as the 2d Ising model and the contact process, the situation is very different. Here the entire dynamical range (apart from corrections due to lattice effects) is determined by the same type of scaling. In fact, in homogeneous critical systems a quenched initial state plays the role of a temporal boundary condition analogous to a surface in a magnetic system at criticality. Such a boundary condition induces one additional surface exponent (here $\lambda_R$) but it neither generates additional scales nor pushes the system into a different universality class. Instead of two dynamically different regimes homogeneous critical systems display a simple crossover phenomenon which can be described in the framework of standard scaling theory. 
\vspace{5mm}

\noindent
\textbf{Acknowledgement:}\\
I thank Malte Henkel for critical reading of the manuscript and useful remarks.
\vspace{5mm}


\noindent{\bf References}
\bibliographystyle{unsrt}
\bibliography{/home/hinrichsen/Dateien/Literatur/master}

\begin{thebibliography}{10}

\bibitem{Young98}
A.~P. Young, editor.
\newblock {\em {Spin Glasses and Random Fields}}.
\newblock World Scientific, New York, 1998.

\bibitem{ZippoldEtAl00}
W.~Zippold, R.~K{\"u}hn, and H.~Horner.
\newblock {Non-equilibrium dynamics of simple spherical spin models}.
\newblock {\em Eur. Phys. J. B}, 13, 2000.

\bibitem{HenkelPleimling03}
M.~Henkel and M.~Pleimling.
\newblock {Local scale invariance as dynamical space-time symmetry in
  phase-ordering kinetics }.
\newblock {\em Phys. Rev. E}, (68):065101, 2003.

\bibitem{McCoyWu73}
B.~M. McCoy and T.~T. Wu.
\newblock {\em {The two-dimensional Ising model}}.
\newblock Harvard University Press, Cambridge, Massachusetts, 1973.

\bibitem{Hinrichsen00}
H.~Hinrichsen.
\newblock {Non-equilibrium critical phenomena and phase transitions into
  absorbing states}.
\newblock {\em Adv. Phys.}, 49:815, 2000.
\newblock [cond-mat/0001070].

\bibitem{Odor04}
G.~\'Odor.
\newblock {Universality classes in nonequilibrium lattice systems}.
\newblock {\em Rev. Mod. Phys.}, 76:663, 2004.

\bibitem{Lubeck04}
S.~L{\"u}beck.
\newblock {Universal scaling behavior of non-equilibrium phase transitions}.
\newblock {\em Int. J. Mod. Phys. B}, 18:3977, 2004.

\bibitem{Henkel07}
M.~Henkel.
\newblock {Ageing, dynamical scaling and its extensions in many-particle
  systems without detailed balance}.
\newblock {\em J. Cond. Mat.}, 19:065101, 2007.

\bibitem{CalabreseGambassi02}
P.~Calabrese and A.~Gambassi.
\newblock {Aging in ferromagnetic systems at criticality near four dimensions}.
\newblock {\em Phys. Rev. E}, 65:066120, 2002.

\bibitem{Hinrichsen06}
H.~Hinrichsen.
\newblock {Is local scale invariance a generic property of ageing phenomena ?}
\newblock {\em J. Stat. Mech.}, page L06001, 2006.

\bibitem{LippielloEtAl06}
E.~Lippiello, F.~Corberi, and M.~Zanetti.
\newblock {Test of Local Scale Invariance from the direct measurement of the
  response function in the Ising model quenched to and to below $T_c$}.
\newblock {\em Phys. Rev. E}, 74:041113, 2006.

\end{thebibliography}

\end{document}